\documentclass[reprint,superscriptaddress,amssymb,amsmath,aps,prx,longbibliography]{revtex4-1}

\usepackage{gensymb}
\usepackage{tikz}
\usepackage[utf8]{inputenc}
\usepackage{graphicx}
\usepackage{amssymb}
\usepackage{epstopdf}
\usepackage{upgreek}
\usepackage{dcolumn}
\usepackage{bm}
\usepackage{braket}
\usepackage{amsmath}
\usepackage{setspace}
\usepackage{mathtools}
\usepackage{float}
\usepackage{tabularx}
\usepackage{tikz}
\usepackage{siunitx}

\expandafter\ifx\csname package@font\endcsname\relax\else
\expandafter\expandafter
\expandafter\usepackage
\expandafter\expandafter
\expandafter{\csname package@font\endcsname}%
\fi

\newcommand{\micro}{$\upmu$}

\newcommand{\be}{\begin{eqnarray}}
\newcommand{\ee}{\end{eqnarray}}
\newcommand{\bfig}{\begin{figure}}
	\newcommand{\efig}{\end{figure}}

\DeclareFontFamily{U}{mathb}{}
\DeclareFontShape{U}{mathb}{m}{n}{
	<-5.5> mathb5
	<5.5-6.5> mathb6
	<6.5-7.5> mathb7
	<7.5-8.5> mathb8
	<8.5-9.5> mathb9
	<9.5-11.5> mathb10
	<11.5-> mathbb12
}{}

\usepackage[colorlinks=true,allcolors=blue]{hyperref}%

\renewcommand{\arraystretch}{1.25} % between columns

\begin{document}

\title{A quantum-enhanced search for dark matter axions}
% \author{The HAYSTAC Collaboration - names will be listed}
\author{K. M. Backes}
\thanks{These two authors contributed equally}
\email{kelly.backes@yale.edu}
\affiliation{Department of Physics, Yale University, New Haven, Connecticut 06511, USA}

\author{D. A. Palken}
\thanks{These two authors contributed equally}
\email{kelly.backes@yale.edu}
\affiliation{JILA, National Institute of Standards and Technology and the University of Colorado, Boulder, Colorado 80309, USA}
\affiliation{Department of Physics, University of Colorado, Boulder, Colorado 80309, USA}

\author{S. Al Kenany}
\affiliation{Department of Nuclear Engineering, University of California Berkeley, California 94720, USA}

\author{B. M. Brubaker}
\affiliation{JILA, National Institute of Standards and Technology and the University of Colorado, Boulder, Colorado 80309, USA}
\affiliation{Department of Physics, University of Colorado, Boulder, Colorado 80309, USA}

\author{S. B. Cahn}
\affiliation{Department of Physics, Yale University, New Haven, Connecticut 06511, USA}

\author{A. Droster}
\affiliation{Department of Nuclear Engineering, University of California Berkeley, California 94720, USA}

\author{Gene C. Hilton}
\affiliation{National Institute of Standards and Technology, Boulder, Colorado 80305, USA}

\author{Sumita Ghosh}
\affiliation{Department of Physics, Yale University, New Haven, Connecticut 06511, USA}

\author{H. Jackson}
\affiliation{Department of Nuclear Engineering, University of California Berkeley, California 94720, USA}

\author{S. K. Lamoreaux}
\affiliation{Department of Physics, Yale University, New Haven, Connecticut 06511, USA}

\author{A.F. Leder}
\affiliation{Department of Nuclear Engineering, University of California Berkeley, California 94720, USA} 

\author{K. W. Lehnert}
\affiliation{JILA, National Institute of Standards and Technology and the University of Colorado, Boulder, Colorado 80309, USA}
\affiliation{Department of Physics, University of Colorado, Boulder, Colorado 80309, USA}
\affiliation{National Institute of Standards and Technology, Boulder, Colorado 80305, USA}

\author{S. M. Lewis}
\affiliation{Department of Nuclear Engineering, University of California Berkeley, California 94720, USA}

\author{M. Malnou}
\affiliation{JILA, National Institute of Standards and Technology and the University of Colorado, Boulder, Colorado 80309, USA}
\affiliation{National Institute of Standards and Technology, Boulder, Colorado 80305, USA}

\author{R. H. Maruyama}
\affiliation{Department of Physics, Yale University, New Haven, Connecticut 06511, USA}

\author{N. M. Rapidis}
\affiliation{Department of Nuclear Engineering, University of California Berkeley, California 94720, USA}

\author{M. Simanovskaia}
\affiliation{Department of Nuclear Engineering, University of California Berkeley, California 94720, USA}

\author{Sukhman Singh}
\affiliation{Department of Physics, Yale University, New Haven, Connecticut 06511, USA}

\author{D. H. Speller}
\affiliation{Department of Physics, Yale University, New Haven, Connecticut 06511, USA}

\author{I. Urdinaran}
\affiliation{Department of Nuclear Engineering, University of California Berkeley, California 94720, USA}

\author{Leila R. Vale}
	\affiliation{National Institute of Standards and Technology, Boulder, Colorado 80305, USA}

\author{E. C. van Assendelft}
\affiliation{Department of Physics, Yale University, New Haven, Connecticut 06511, USA}

\author{K. van Bibber}
\affiliation{Department of Nuclear Engineering, University of California Berkeley, California 94720, USA}

\author{H. Wang}
\affiliation{Department of Physics, Yale University, New Haven, Connecticut 06511, USA}

\date{\today}

%\section{Intro paragraph}
%\begin{multicols}{1}
\begin{abstract}
In dark matter axion searches, quantum uncertainty manifests as a fundamental noise source, limiting the measurement of the quadrature observables used for detection. We use vacuum squeezing to circumvent the quantum limit in a search for a new particle. By preparing a microwave-frequency electromagnetic field in a squeezed state and near-noiselessly reading out only the squeezed quadrature \cite{malnou2019squeezed}, we double the search rate for axions over a mass range favored by recent theoretical projections \cite{buschmann2020sims,klaer2017mass26}. We observe no signature of dark matter axions in the combined 16.96--17.12 and 17.14--17.28\,$\upmu\text{eV}/c^2$ mass window for axion-photon couplings above $g_{\gamma} = 1.38\times g_{\gamma}^\text{KSVZ}$, reporting exclusion at the 90\% level. 
\end{abstract}

\maketitle
\section*{Introduction}

% \cite{peccei1977CP, wilczek1978PT, weingerg1978boson}
The manipulation of quantum states of light \cite{slusher1985observation} has long held the potential to enhance searches for new fundamental physics. Only recently has the maturation of quantum squeezing technology coincided with the emergence of fundamental physics searches limited by quantum uncertainty \cite{tse2019quantum, brubaker2017first}. In particular, the QCD axion, originally ``invented" to solve the strong charge-parity (${CP}$) problem of quantum chromodynamics \cite{peccei1977CP}, may constitute the dark matter which makes up 27\,\% of the universe's energy density \cite{preskill1983cosmology, dine1983harmless, Abbott1983bound, ade2016planck}. Today, roughly a century after it was first postulated, dark matter remains one of the most profound mysteries in fundamental physics. It determines cosmic structure formation and dominates the dynamics of galaxies, and there is overwhelming evidence that it cannot be composed of any particles described by the Standard Model of particle physics. 

Axions have in recent years emerged as leading dark matter candidates. The case for dark matter axions has been bolstered as the concepts and technologies used to attempt to detect them have developed \cite{ouellet2019first, majorovits2020madmax, Geraci2014ARIADNE, Garcon2017CASPr, malnou2018optimal,malnou2019squeezed, brubaker2017first}, and experimental null results have placed stringent constraints on prominent alternatives \cite{Bertone2018WIMP}. If they exist, axions would likely be many orders of magnitude lighter than all massive Standard Model particles. In fact, they are sufficiently low-energy as to behave like a weakly coupled oscillating field permeating all space. Investigating this field's existence requires detectors sensitive to the axion's coherent effects, rather than single-particle interactions. Recent years have seen a proliferation of such detector platforms, capable of probing different possible values of the axion mass \cite{ouellet2019first, majorovits2020madmax, Geraci2014ARIADNE, Garcon2017CASPr}. Among these, axion haloscopes \cite{brubaker2017first, zhong2018results, braine2020extended_ADMX, lee2020CAPP}, designed to search within the $1$--$50\,\upmu\text{eV}/c^2$ mass range, are, to date, the only platforms to demonstrate sensitivity to the QCD axion.

An axion haloscope exploits the coupling $g_\gamma$ of the hypothetical axion field $a$ to the pseudoscalar electromagnetic field product $\textbf{E}\cdot\textbf{B}$. It comprises a tunable, high-quality factor ($Q$) cavity embedded in a large static magnetic field $\textbf{B}$ \cite{sikivie1983haloscope}, coupled to a low-noise readout system, and held at a cryogenic temperature. An oscillating axion field generates a feeble oscillating electric field $\textbf{E}$ which is resonantly enhanced when the frequency corresponding to the axion's mass, $\nu_a = m_a c^2/h$, falls within the bandwidth of a $\text{TM}_{0n0}$-like mode of the cavity. The critical feature of axion searches is achieving sensitivity over a broad frequency range to detect a narrowband, axion-induced power excess of order $10^{-23}$\,W at the unknown frequency $\nu_a$.

\begin{figure*}[t] % full width figure
    \centering
    \includegraphics[width=0.9\textwidth]{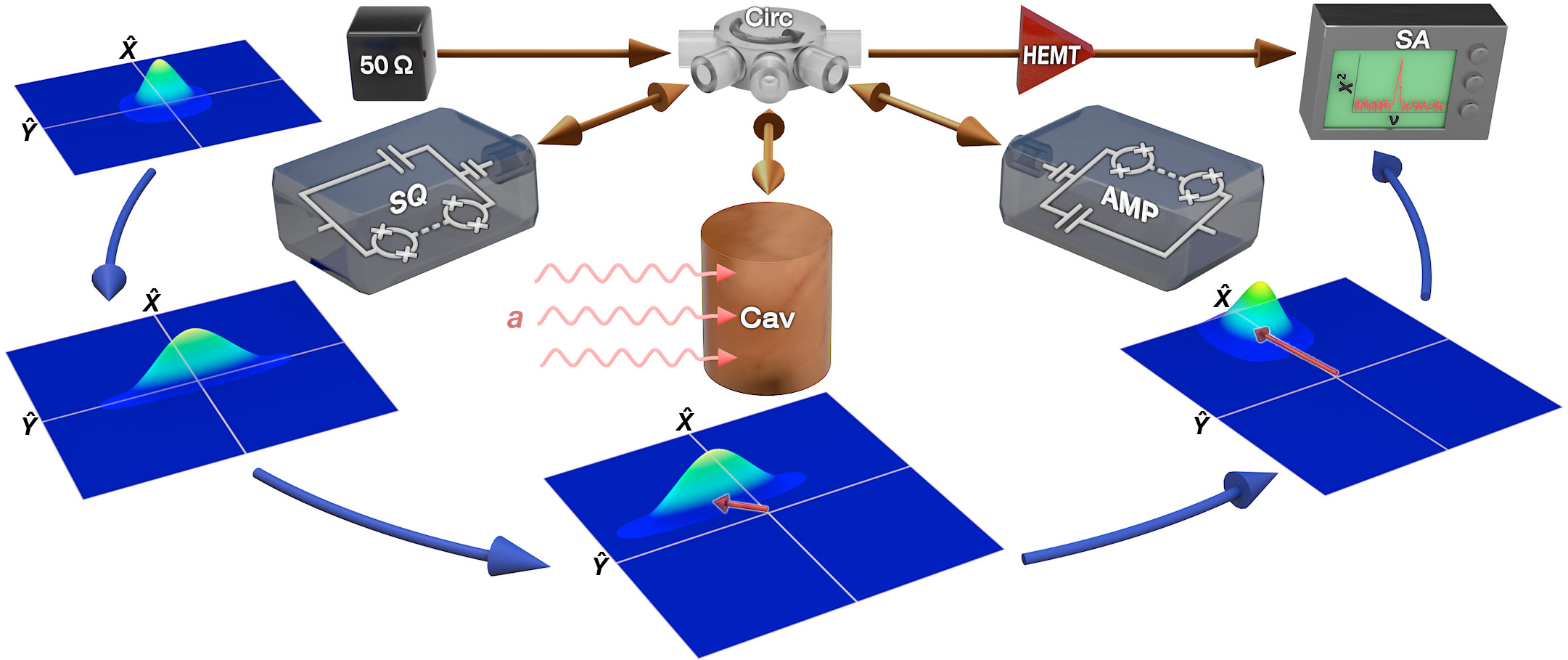}
    \caption{Illustration of the squeezed state receiver-equipped haloscope showing the transformation of the vacuum state in quadrature space. A vacuum state, whose Wigner function (color surface) \cite{Braunstein2005wigner} is symmetric in quadratures $\hat{X}$ and $\hat{Y}$, is sourced as Johnson-Nyquist noise from a \SI{50}{\ohm} microwave termination (black box) at 61\,mK. It is routed by a nonreciprocal element (Circ) to the SQ Josephson parametric amplifier (JPA) which squeezes the $\hat{X}$ quadrature. The squeezed state may then be displaced by a hypothetical axion field $a$ in the axion cavity (Cav). It is subsequently unsqueezed by the AMP JPA, which in the process amplifies the axion-induced displacement along $\hat{X}$. The resulting state is measured by a conventional microwave receiver led by a high-electron-mobility transistor (HEMT) amplifier. The time record of many realizations of this process are Fourier transformed for subsequent spectral analysis (SA).}
    \label{fig:SSR}
\end{figure*}

The main figure of merit for a haloscope search is thus the scan rate $R$ (in Hz/s) at which the detector can tune through frequency space for a given ratio of axion signal to noise power. Axion signal power in a haloscope is determined by the magnetic field, cavity performance, and physical parameters of the axion field. With the signal power fixed by those quantities, a given haloscope has a quantum-limited scan rate imposed by the zero-point fluctuations of the axion-sensitive cavity mode \cite{malnou2019squeezed}. These fluctuations create a fundamental barrier to improving haloscope scan rates that can only be overcome using quantum-enhanced measurements \cite{CavesThorne1982oscillator}. Without a means of bypassing the quantum limit, the highly unfavorable $R \propto \nu^{-14/3}$ frequency scaling \cite{palken2020improved} of single-cavity haloscopes poses a stark challenge. At the quantum limit, scanning the $1$--$10$\,GHz frequency decade at the benchmark KSVZ \cite{kim1979KSVZ, shifman1980KSVZ2} coupling would require hundreds of years of live time for today's state-of-the-art haloscopes. Few axion dark matter searches have approached this limit \cite{brubaker2017first, braine2020extended_ADMX}, and until now none have exceeded it. Novel measurement technology is therefore needed to circumvent the quantum limit. 
%  and continue the trend of decreasing haloscope noise levels which has been predominantly responsible for improving scan rates since 1987 \cite{depanfilis1987limits_RBF, hagmann1998results_ADMX, asztalos2010squid_ADMX, brubaker2017first}

Here we report on a quantum-enhanced axion search. The search was carried out with the Haloscope At Yale Sensitive To Axion Cold dark matter (HAYSTAC), designed to detect QCD axions in the $m_a > 10\,\upmu\text{eV}/c^2$ range favored by large-scale lattice QCD calculations of post-inflationary Peccei-Quinn symmetry breaking scenarios \cite{buschmann2020sims, klaer2017mass26}. This work surpasses the quantum limit by coupling the HAYSTAC cavity to the squeezed state receiver (SSR) \cite{malnou2019squeezed} shown in Fig.\,\ref{fig:SSR}. The SSR comprises two flux-pumped Josephson parametric amplifiers (JPAs) \cite{yamamoto2008flux} coupled to the axion cavity via a microwave circulator. The first JPA, labeled the ``squeezer" (SQ), prepares a squeezed vacuum state, which is coupled into the axion cavity and subsequently measured noiselessly using the second, ``amplifier" (AMP) JPA. We achieve 4.0\,dB of off-resonant vacuum squeezing, \textit{after} the state is degraded by transmission losses and added noise, yielding a factor-of-1.9 scan rate enhancement beyond what would have been achievable at the quantum limit. Breaking through the quantum limit invites a new era of fundamental physics searches in which noise reduction techniques yield unbounded benefit rather than the diminishing returns of approaching the quantum limit.

%\newpage
%The central innovation of Phase II of the HAYSTAC experiment is the use of squeezed noise to improve the spectral scan rate for dark matter axions. 
\section*{Squeezed State Receiver}
The SSR \cite{malnou2019squeezed} is coupled to the cavity, which is governed by the Hamiltonian
\begin{equation}
\hat{H} = \frac{h\nu_c}{2}\left(\hat{X}^2+\hat{Y}^2\right), 
\end{equation}
where $\hat{X}$ and $\hat{Y}$ are the cavity field's quadratures and obey $[\hat X, \hat Y] = i$. The SQ, with half-pump frequency centered on the cavity frequency, $\nu_p/2 = \nu_c$, performs a unitary squeezing operation on the electromagnetic field at its input, reducing the $\hat X$ quadrature variance below vacuum and amplifying the $\hat Y$ quadrature variance in accordance with the uncertainty principle. An axion field, if present, displaces the squeezed state from the origin in the cavity field quadrature phase space, proportional to the Lorentzian transmission profile of the cavity at $\nu_a$. The squeezed quadrature is then amplified by the AMP via the inverse operation used to perform the squeezing. In the absence of any loss the entire process is noiseless and unitary.

%There is loss in the cavity and in transporting the squeezed state to the amp. There is loss. it is kinda inevitable. In particular, because the cavity has loss. In the presence of cavity loss (3:40 PM).  squeezing improves the bandwidth
Squeezing improves the bandwidth over which the apparatus is sensitive to an axion, rather than its peak sensitivity, obtained at the cavity resonant frequency \cite{malnou2019squeezed}. This bandwidth increase can be understood by considering the behavior of the three distinct noise sources within a haloscope which obscure an axion-induced signal. The first is Johnson-Nyquist noise (which in the zero-temperature limit is vacuum noise) $N_c$ sourced from the internal loss of the cavity as a consequence of the quantum fluctuation-dissipation theorem. This noise is by definition inaccessible to the experimentalist, and cannot be squeezed. It is filtered by the cavity response, giving it a Lorentzian profile at the cavity output. The second noise source is the system added noise $N_A$, which encompasses the added noise of the entire amplification chain, including the AMP, referred to the input of the AMP. This noise source has historically been a dominant or co-dominant contribution \cite{lee2020CAPP, braine2020extended_ADMX, zhong2018results}, but here we operate our JPAs in a phase-sensitive mode, which can completely eliminate the contribution of this noise in one quadrature \cite{caves1982quantum, malnou2018optimal} (switching from phase-insensitive to phase-sensitive operation on its own does \textit{not} improve scan rate, but only phase-sensitive operation allows for a benefit from squeezing; see Appendix C of Ref.\,\cite{malnou2019squeezed}). In Appendix \ref{app:EO}, we show that our added noise is sufficiently small as to be of negligible importance at all frequencies of interest. The third noise source is Johnson-Nyquist noise $N_r$ incident on and reflected off the cavity, which dominates away from cavity resonance. This noise is invariably present in any receiver configuration, and in our setup it is sourced from a \SI{50}{\ohm} termination held at the cryostat base temperature (\SI{61}{\milli\kelvin}). The ratio of the signal spectral density $S_\text{ax}(\delta_\nu)$ that would be delivered by an axion at any given detuning $\delta_\nu = \nu - \nu_c$ from cavity resonance to the cavity noise $N_c(\delta_\nu)$ is spectrally constant. Thus, the axion visibility $\alpha(\delta_\nu)$, defined as the signal-to-noise power spectral density (PSD) ratio (neglecting transmission losses)
\begin{equation}\label{vis} % equation for visibility. we should format this to try and match our previous papers 
    \alpha(\delta_\nu) =  \frac{S_\text{ax}(\delta_\nu)}{N_c(\delta_\nu)+N_A(\delta_\nu)+N_r(\delta_\nu)}
\end{equation}
is maximized on resonance where $N_c$ is most dominant over $N_r$. Here, the denominator is the noise power spectral density comprised of the sum of the three noise sources listed above. The maximum visibility $\alpha(0)$ occurs for an on-resonance axion signal. However, the scan rate $R$ also depends on the bandwidth over which high visibility is maintained \cite{malnou2019squeezed}:
\begin{equation}
 R\propto\int_{-\infty}^\infty\alpha^2(\delta_\nu) d\delta_\nu.
\end{equation}
By squeezing $N_r$, the SSR does not improve the maximal visibility $\alpha(0)$, but rather increases the frequency range over which $N_c$ is the dominant noise source, improving $R$ and increasing the number of potential axion masses simultaneously probed at each cavity tuning step. 

\begin{figure}[!ht] 
    \centering
    \includegraphics[width=\columnwidth]{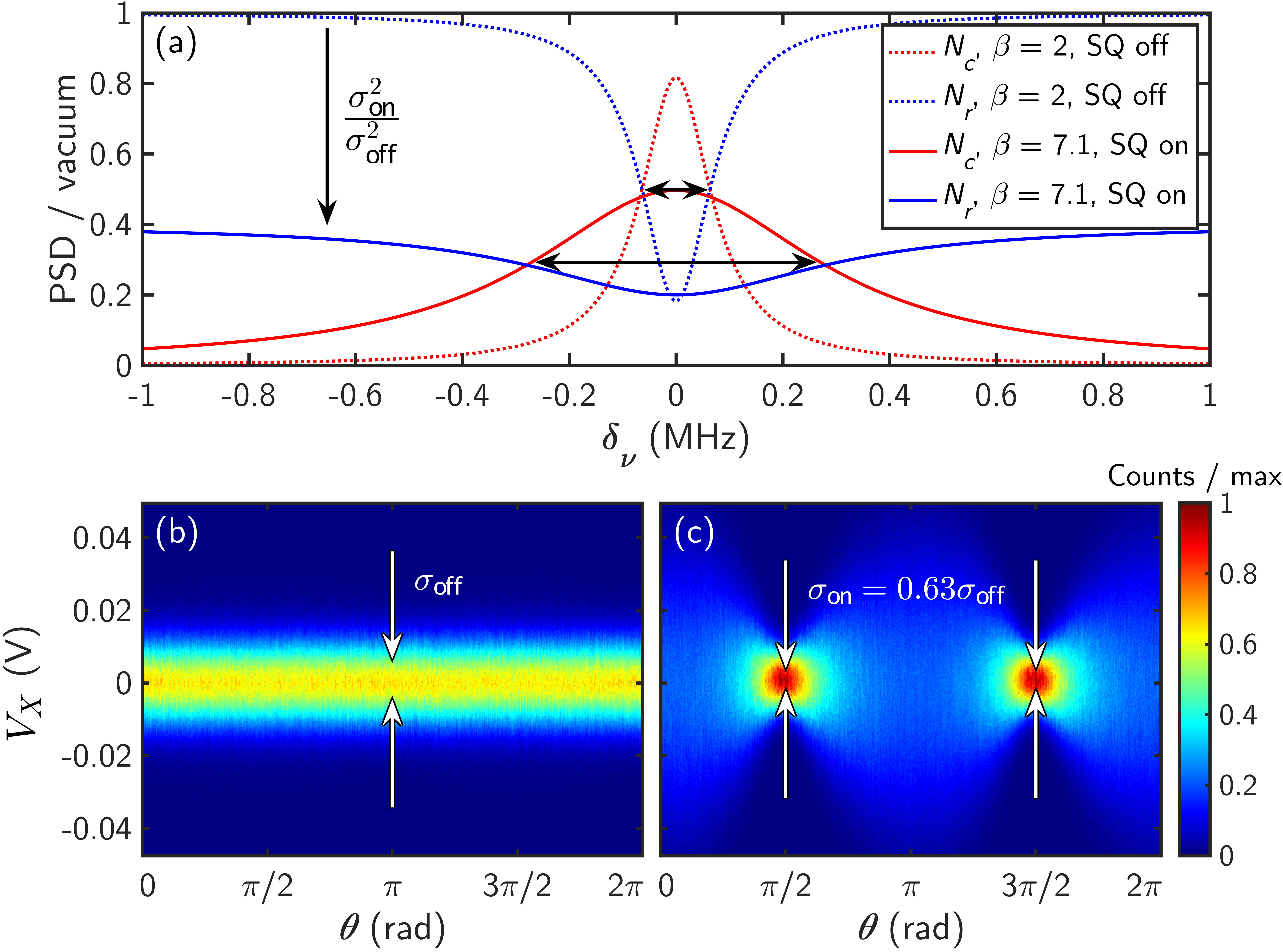}
    \caption{Advantage conferred by squeezing. (a) Theory curves show power spectral density (PSD) of Johnson-Nyquist noise reflected off the cavity ($N_r$, blue) and generated within it ($N_c$, red) at the cavity output, as a function of frequency. Solid lines indicate operating parameters in this work and dashed lines indicate the coupling configuration used in previous HAYSTAC operation \cite{brubaker2017first}. The combination of squeezing and larger cavity coupling increases the bandwidth over which HAYSTAC is sensitive to axions. (b) and (c) Voltage fluctuation ($V_X$) histograms as a function of the phase $\theta$ between the squeezer (SQ) and amplifier (AMP) pumps. In (b), the SQ is turned off such that it reflects the vacuum noise untransformed, and the measured variance $\sigma_\mathrm{off}^2$ is independent of phase. (c) The SQ is turned on and the variance $\sigma_\text{on}^2$ is minimized to a value $ < \sigma_\text{off}^2$ for $\theta = \pi/2$ and $3\pi/2$.}
    \label{fig:sqdata}
\end{figure}

Because squeezing reduces $N_r$, it is then beneficial to overcouple the cavity's measurement port relative to its internal loss rate in order to increase the measurement bandwidth \cite{malnou2019squeezed}. The coupling ratio $\beta = \kappa_m/\kappa_l$, where $\kappa_l$ is the internal dissipation rate and $\kappa_m$ is the rate at which the cavity state decays out the measurement port, should be set to approximately twice the deliverable squeezing, defined as the variance reduction at the system output $S = \sigma_\text{on}^2/\sigma_\text{off}^2$. In Fig.\,\ref{fig:sqdata}a, the dashed lines show $N_c(\delta_\nu)$ in red and $N_r(\delta_\nu)$ in blue with $\beta = 2$ \cite{brubaker2017first}, the ideal coupling ratio for a quantum-limited haloscope. The cavity noise $N_c$, emerging from the cavity in $\delta_\nu$-independent ratio with a hypothetical axion signal, is only dominant over the narrow band shown by the upper black arrow. The solid lines show the operating conditions for this work, where squeezing is employed and the cavity is overcoupled at $\beta = 7.1$ (slightly higher than 2$S$ due to excess thermal noise; see Appendix \ref{app:EO}). Squeezing reduces the reflected noise, and increasing $\beta$ to 7.1 increases the cavity bandwidth. Without squeezing, $\beta = 7.1$ would make cavity-reflected noise dominant at all frequencies. Squeezing this noise below the quantum limit yields high visibility extended far off resonance. 
% \footnote{The overcoupling here is slightly higher than twice the delivered squeezing $S$ because the cavity thermal noise is higher than the temperature of the cryostat (see Methods section). When not squeezing, the optimal overcoupling is actually $\beta \approx 2.8$, slightly higher than the value $\beta = 2$ plotted in Fig.\,\ref{fig:sqdata}. Fully accounted for, the cavity thermal noise has negligible affect on the scan rate enhancement achieved via squeezing.}

Figure \ref{fig:sqdata}b and c show a direct measurement of the reduction in variance relative to vacuum in the AMP's amplified $(\hat{X})$ quadrature, as a function of the phase $\theta$ between SQ and AMP pumps. Noise is minimized when $\theta$ is an odd multiple of $\pi/2$, such that the AMP amplifies the quadrature that was initially squeezed. We automatically stabilize the phase by using a single microwave generator to source both pump tones. Comparing amplified vacuum with the SQ on versus off, we measure a delivered squeezing of $S = 0.40$ (equivalently, $-10\log10(S) = 4.0$\,dB) off cavity resonance. Squeezing of $N_r$ is benchmarked off resonance where it is the dominant noise. In practice, squeezing is limited by $\eta$, the transmissivity of the cables and microwave components between the SQ and AMP, scaling as \(S = \eta G_s+(1-\eta)\) where $G_s$ is the inferred squeezing at the output of the squeezer, limited by saturation effects \cite{malnou2018optimal}. In the HAYSTAC system, $\eta \approx 0.63$, giving $S = 0.37$ as the theoretical maximum squeezing. This is nearly saturated by the measured value and corresponds to an almost twofold scan rate enhancement relative to optimal unsqueezed operation. 

\section*{Results}

Using the SSR apparatus described above, we probed over 70\,MHz of well-motivated parameter space \cite{klaer2017mass26, buschmann2020sims} in half the time that would have been required for unsqueezed operation, saving approximately 100 days of scanning. Initial data acquisition occurred from September 3 to December 17, 2019, covering $4.100$--$4.178$\,GHz and skipping a TE mode at $4.140$--$4.145$\,GHz which does not couple to the axion \cite{Rapidis2019cavity}. A total of 861 spectra were collected, of which 33 were cut due to cavity frequency drift, poor JPA performance, or an anomalous power measurement in a probe tone injected near cavity resonance. Analysis of the initial scan data yielded 32 power excesses that merited further scanning, consistent with statistical expectations \cite{brubaker2017analysis}. Rescan data were collected from February 25 to April 11, 2020.  None of the power excesses from the initial scan persisted in the analysis of rescan data. The process of optimizing the SSR as well as further measurements and calibrations taken periodically are described in Appendix \ref{app:EO}. 

\begin{figure}[!ht] % full width figure    
%\centering
    \includegraphics[width=\columnwidth]{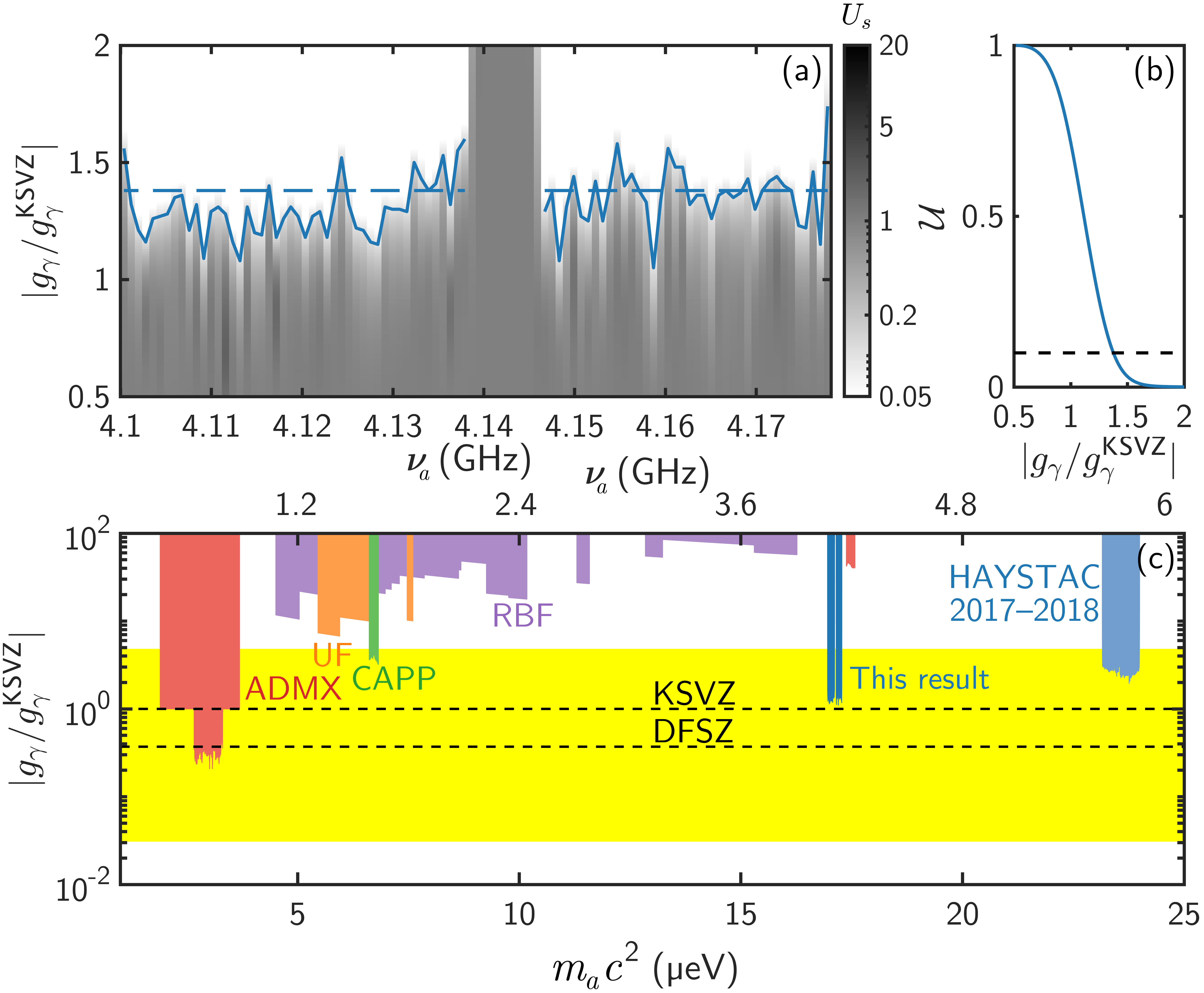}
    \caption{Axion exclusion from this work. (a) Prior updates $U_s$ in greyscale in the two-dimensional parameter space of axion frequency $\nu_a$ and coupling $g_\gamma$ are achieved with a Bayesian analysis framework \cite{palken2020improved}. The 10\% prior update contour is shown in solid blue. The corresponding $90\%$ aggregate exclusion level of $1.38\,\times\,g_\gamma^\text{KSVZ}$ is shown as dashed blue. (b) The frequency-resolved $U_s$ are combined into a single aggregate prior update $\mathcal{U}$ as a function of coupling $g_\gamma$ over the entire frequency range covered by the dashed blue line in (a). (c) Results of this work are shown alongside previous exclusion results from other haloscopes (see Ref.\,\cite{Tanabashi2018PDB}). The QCD axion model band \cite{cheng1995axion} is shown in yellow, with the specific KSVZ \cite{kim1979KSVZ, shifman1980KSVZ2} and DFSZ \cite{dine1981DFSZ, zhit1980DFSZ2} model lines shown as black dashed lines.}%haloscopes shown in red (ADMX) \cite{hagmann1998results_ADMX, asztalos2002experimental_ADMX, asztalos2004improved_ADMX, asztalos2010squid_ADMX, sloan2016limits_ADMX, du2018search_ADMX, braine2020extended_ADMX, boutan2018sidecar_ADMX}, orange \cite{hagmann1990search_UF, hagmann1990results_UF}, purple \cite{wuensch1989results_RBF, depanfilis1987limits_RBF}, and green \cite{lee2020CAPP}. The QCD axion model band is shown in yellow \cite{cheng1995axion}, with the specific  KSVZ \cite{kim1979KSVZ, shifman1980KSVZ2} and DFSZ \cite{dine1981DFSZ, zhit1980DFSZ2} model lines shown as black dashed lines.}
    \label{fig:EXC}
\end{figure}

From this data, we report a constraint on axion masses $m_a$ within the 16.96--17.12 and 17.14--17.28\,$\si{\micro\electronvolt}/c^2$ windows. Using the Bayesian power-measured analysis framework \cite{palken2020improved} described in Appendix \ref{app:DA}, we exclude axions with $g_\gamma\geq 1.38\times g_\gamma^\text{KSVZ}$. Figure \ref{fig:EXC}a shows in greyscale the prior update (change in probability) $U_s$ that the axion resides at any specific location in parameter space, with the solid blue line showing the coupling where $U_s = 10\%$ at each frequency. The aggregate update $\mathcal{U}$ (blue curve in Fig.\,\ref{fig:EXC}b) to the relative probabilities of the axion and no-axion hypotheses corresponds to exclusion at the 90\% confidence level over the entire window at the coupling where $\mathcal{U} = 10\%$. The results from our quantum-enhanced data run are shown alongside other axion haloscope exclusion curves in Fig.\,\ref{fig:EXC}c, including previous HAYSTAC results \cite{brubaker2017first, zhong2018results}, which operated a single JPA near the quantum limit. 
% Whereas previous HAYSTAC data runs \cite{brubaker2017first, zhong2018results} operated a single current-pumped Josephson parametric amplifier (JPA) \cite{castellanos2007widely} near the quantum limit,

\section*{Conclusion} 
%\label{Sec:Conclusion}
With these results, the HAYSTAC experiment has achieved a breakthrough in sensitivity by conducting a sub-quantum-limited search for new fundamental particles. Through the use of a squeezed state receiver which delivers 4.0\,dB of off-resonant noise variance reduction relative to vacuum, we have demonstrated record sensitivity to axion dark matter in the $10$\,$\upmu\text{eV}/c^2$ mass decade. This work demonstrates that the incompatibility between delicate quantum technology and the harsh and constrained environment of a real search for new particle physics can be overcome: in this instance in an axion haloscope requiring efficient tunability and operation in an 8\,T magnetic field. As intense interest in quantum information processing technology continues to drive transmission losses downward, quantum-enhanced measurement will deliver transformative benefits to searches for new physics. In particular, the prospect of removing nonreciprocal signal routing elements \cite{burkhart2020error} would boost transmission efficiencies above 90\%, yielding a greater-than-tenfold scan rate increase beyond the quantum limit \cite{malnou2019squeezed}.

\section{Acknowledgements}
We acknowledge support from the National Science Foundation under grant numbers PHY-1701396, PHY-1607223, PHY-1734006 and PHY-1914199 and the Heising-Simons Foundation under grants 2014-0904 and 2016-044. We thank Kyle Thatcher and Calvin Schwadron for work on the design and fabrication of the SSR mechanical components, Felix Vietmeyer for his work on the room temperature electronics, and Steven Burrows for graphical design work. We thank Vincent Bernardo and the J. W. Gibbs Professional Shop as well as Craig Miller and Dave Johnson for their assistance with fabricating the system's mechanical components. We thank Dr. Matthias Buehler of low-T Solutions for cryogenics advice. Finally, we thank the Wright laboratory for housing the experiment and providing computing and facilities support.

\appendix

\section{EXPERIMENTAL OPERATIONS}\label{app:EO}
\begin{figure*}[t] % full width figure    
    \centering
    \includegraphics[width=1\textwidth]{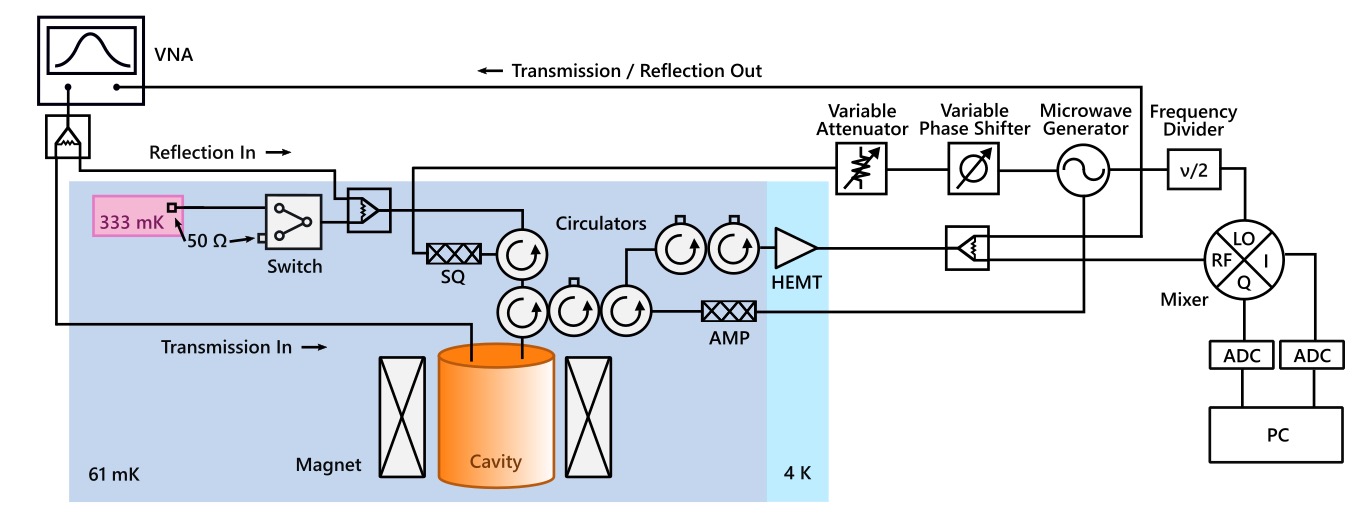}
    \caption{Simplified HAYSTAC experimental diagram. A single signal generator provides the local oscillator (LO) tone as well as the tones for pumping both JPAs (box with three x's). Each JPA has two ports: one for the input of probe tones and one where the signal is input/output. The LO is set at half the pump frequencies via a frequency divider, and the relative phase and amplitude of the pump tones are set using a variable phase shifter and attenuator on the SQ pump line. Switches in the SQ and AMP pump lines (not shown) are used to toggle the JPAs on and off. Microwave circulators route signals nonreciprocally in order to realize the time-sequence of operations illustrated in Fig.\,\ref{fig:SSR}. Circulators with a \SI{50}{\ohm} termination on one port act as isolators, shielding upstream circuit elements from unwanted noise coming from further down the measurement chain. During data acquisition and calibration measurements, signal and noise emitted from and reflected off the cavity are amplified by a HEMT amplifier at 4\,K, fed into the RF port of an IQ mixer and mixed down to an intermediate frequency, digitized (ADC), and read into the computer (PC) where the PSD is calculated. The cavity's Lorentzian profile is monitored with reflection and transmission measurements taken using a vector network analyzer (VNA), for which a portion of the output is split off before the mixer. A switch that toggles between a hot (333\,mK) and cold (61\,mK) \SI{50}{\ohm} loads is used for calibration measurements described in the text.}
    \label{fig:SCHEM}
\end{figure*}

The HAYSTAC experiment, shown schematically in Fig.\,\ref{fig:SCHEM}, takes place in an LD250 BlueFors dilution refrigerator (with base temperature $61$\,mK) attached to an 8\,T solenoidal magnet with a counterwound bucking coil. The cavity sits at the center of the high-magnetic field region, has volume $V_C = 1.5$\,L, unloaded quality factor $Q_0 = 2\pi\nu_c/\kappa_l = 47000 \pm 5000$, and TM$_{010}$ mode form factor $C_{010}\approx0.5$. The cavity has two antenna ports \cite{Kenany2017design}. The first is a weakly coupled port used to input microwave tones to monitor gain stability and measure the cavity transmission profile. The second is a strongly coupled port which is used to read out from the cavity and for VNA scattering parameter measurements. The coupling of the coaxial antenna is mechanically actuated with a stepper motor. The two JPAs that comprise the SSR are held above the cavity inside a four-layer shielding can (niobium, Amumetal 4K, aluminum, Amumetal 4K; listed from inside to outside), surrounded by three superconducting bucking coils. This shielding is sufficient to reduce the flux through each of the JPAs' SQUID loops to much less than one magnetic flux quantum, a roughly millionfold decrease in field strength from the high-field region 1\,m below. Inside the shielding can, each JPA has a current bias coil used to tune its resonant frequency. We tune the cavity by rotating an off-axis copper-plated tuning rod which occupies approximately one quarter of the cavity's volume using an Attocube ANR240 piezoelectric motor.

The experiment is operated in a series of discrete tuning steps. At each, we actuate the piezo to tune the TM$_{010}$-like cavity mode, then extract the cavity resonance $\nu_c$, loaded quality factor $Q_L = 2\pi\nu_c/(\kappa_m+\kappa_l)$, and coupling factor $\beta$ using pairs of VNA measurements taken in transmission and reflection. In order to center the amplification and analysis bands on cavity resonance, we use a single microwave generator and a frequency divider to set the two JPA pump frequencies to $2\nu_c$ and the local oscillator used for homodyne measurement to $\nu_c$. We set the gain of the AMP to $G_A \approx 28$\,dB by adjusting the microwave generator amplitude and bias coil current. Then, the relative pump phase $\theta$ and amplitude are optimized to maximize squeezing using an electronically actuated variable phase shifter and attenuator. Axion-sensitive voltage fluctuations $V_X(t)$ are collected for $\tau=3600$\,s in 720 5\,s chunks. These chunks are further broken down into 10\,ms segments before being Fourier transformed, having their PSDs computed, and being added to a running average. Each power spectrum then encompasses 3600\,s worth of data and is truncated to the analysis band of approximately 3\,MHz centered on the cavity frequency. 
%We restrict all analysis to an IF band of full width 1.3 MHz & 2∆νc centered on the cavity mode at 780 kHz. - wording from the PRL 

To characterize our sensitivity to the axion, we perform calibrations once every nine tuning steps. The first step in our calibration protocol is a set of three power spectrum measurements to determine two experimental parameters: the noise spectral density 
\begin{equation}
N_{c0} = \frac 12 \left(\frac 12 + \frac{1}{e^{h \nu /k_BT} - 1}\right) = \frac{1}{4}\coth\left(\frac{h \nu}{2 k_B T}\right)
\end{equation}
generated within the cavity before being filtered by the cavity transmission profile, and the squeezing $G_s(\nu)$ inferred at the output of the SQ. Measurement 1 is taken off cavity resonance with the squeezer turned off. Measurements 2 and 3 are taken on cavity resonance with the squeezer turned on and off respectively. From measurements 2 and 3 we infer a typical $G_s \approx 0.1$, which corresponds to slightly less than the 4.0\,dB of squeezing obtained in our off-resonant measurement of Fig.\,\ref{fig:sqdata}. From measurements 1 and 3 we calculate $N_{c0} = 0.41\pm 0.02$ quanta, larger than the $N_f = 0.27$ quanta expected in thermal equilibrium with the cryostat base plate, due to imperfect thermalization of the tuning rod. Because of this resonant excess over the noise sourced by the \SI{50}{\ohm} termination at the SQ input, which does equilibrate to the base temperature, the coupling of the strong port has to be set to slightly higher than twice the delivered squeezing $S$. Given this excess cavity noise, the optimal overcoupling for $S = 0.40$ is $\beta = 7.1$ instead of $\beta = 4.5$. Without squeezing, the optimal overcoupling in the presence of excess cavity noise is $\beta \approx 2.8$, slightly higher than the value $\beta = 2$ plotted in Fig.\,\ref{fig:sqdata}. Fully accounted for, the cavity Johnson-Nyquist noise has negligible effect on the scan rate enhancement achieved via squeezing.

The second calibration measurement is a thermal calibration similar to the protocol used in previous HAYSTAC results \cite{malnou2018optimal,brubaker2017first}, with a cold load at the dilution refrigerator base temperature and hot load maintained at 333\,mK (monitored by a Magnicon SQUID-based temperature sensor). This gives us the frequency-dependent, single-quadrature system added noise $N_A(\nu)$ referred to the input of the AMP. In our previous work, phase-insensitive JPA operation required the addition of $N_A\geq1/4$ quanta per quadrature. Now, operating in phase-sensitive mode, we calculate the system noise referred to the AMP input to be $N_A = 0.03\pm0.02$ quanta averaged over the analysis band. These calibrations, together with system loss measurements taken \textit{ex-situ}, provide an accurate measurement of the total noise against which an axion signal must be measured.

\section{DATA ANALYSIS} \label{app:DA}
Our data processing largely follows Ref.\,\cite{brubaker2017analysis}, and our analysis framework is that of Ref.\,\cite{palken2020improved}. The processing and analysis were separately performed by two semi-independent analysis teams at JILA and Yale, ultimately agreeing within 1\% on the couplings excluded. The results presented in Fig.\,\ref{fig:EXC} are the average of the two analyses.

Coarse spectral structure and non-axionic power excesses are separately removed in both the intermediate frequency band (IF, DC--MHz) and the radio frequency band (RF, GHz). After cutting spectra as discussed in the main text, we average the spectra in the IF band in order to extract shared non-axionic structure and power excesses. We then remove remaining spectral structure wider than the axion ($\sim 9$\,kHz) from each individual spectrum and perform cuts only on highly anomalous spikes which do not behave as axions in the RF band. Structure removal in both the IF and RF bands is done by dividing out Savitzky-Golay filters, equivalent to polynomial generalizations of moving averages in the frequency domain \cite{brubaker2017analysis}. The remaining processed spectra consist of Gaussian-distributed noise with mean $\mu = 0$ (in the case that no axion is present) and standard deviation $\sigma=1/\sqrt{\Delta_b \tau} = 0.0017$, where $\Delta_b = 100$\,Hz is the Fourier bin size. Using maximum likelihood weights determined from the noise calibrations of the previous section and the parameters that determine axion signal power, we add these spectra together to produce a single combined spectrum. We then coadd groups of adjacent frequencies, taking into account the virialized axion lineshape, to yield the final grand spectrum. 

To test for the presence of an axion signature in the grand spectrum, we use the Bayesian power-measured analysis framework of Ref.\,\cite{palken2020improved}. The framework constitutes a straightforward application of Bayes' theorem which uses both experimental sensitivity to the axion $\eta_i$ (defined as the signal-to-noise ratio for an axion with coupling strength $g_\gamma = 1$) and actual excess power $x_i$ measured at each frequency bin $i$ to test for the presence or absence of the axion, taking advantage of the full information content of the measurement. 

The single-frequency, single-scan prior updates $u_i$ are defined as the change in probability of the axion existing at the $i^\text{th}$ grand spectrum bin due to the grand spectrum power $x_i$ measured there. This reduces to the ratio of the no-axion $\mathcal{N}_i$ and axion $\mathcal{A}_i$ distributions, well approximated as Gaussian with $\sigma=1$ and $\mu_0=0$ (no-axion) or $\mu_a = g_{\gamma ,i}^2 \eta_i$ (axion), evaluated at $x_i$:
\begin{equation}\label{pup}
    u_i = \frac{P\big(x_i\big|\mathcal{A}_i\big)}{P\big(x_i\big|\mathcal{N}_i\big)} = \exp\bigg[-\frac{\mu_{a,i}^2}{2}+\mu_{a,i}x_i\bigg].
\end{equation}
We multiply the updates from rescans onto the corresponding bins' initial scan updates to get the total updates $U_i$ for each frequency bin. The resulting aggregate prior update,
\begin{equation}\label{agg}
    \mathcal{U}(g_\gamma) = \frac{1}{N}\sum_{i=1}^NU_i(g_\gamma),
\end{equation}
which fully accounts for the look-elsewhere effect, falls below $10\%$ at $g_{\gamma} = 1.38 \times g_{\gamma}^\text{KSVZ}$ for the $4.100$--$4.140$ and $4.145$--$4.178$\,GHz combined frequency range shown in Fig.\,\ref{fig:EXC}. In other words, the probability of an axion existing in our scan window at or above 1.38\,$\times$\,$g_{\gamma}^\text{KSVZ}$ has decreased at least $90\%$ as a result of our measurement. This corresponds to exclusion at the 90\% confidence level in the sense reported in e.g. Ref.\,\cite{zhong2018results} (see Appendix A of Ref.\,\cite{palken2020improved}). Finally, the subaggregated updates $U_s$ shown in Fig.\,\ref{fig:EXC}a apply the aggregation formula, Eq.\,\eqref{agg}, to 100 independent windows each covering 1\% of the width of the exclusion plot.

\bibliographystyle{naturemag} 

\bibliography{biblio.bib}

\newcommand{\noop}[1]{}
\begin{thebibliography}{10}
\expandafter\ifx\csname url\endcsname\relax
  \def\url#1{\texttt{#1}}\fi
\expandafter\ifx\csname urlprefix\endcsname\relax\def\urlprefix{URL }\fi
\providecommand{\bibinfo}[2]{#2}
\providecommand{\eprint}[2][]{\url{#2}}

\bibitem{malnou2019squeezed}
\bibinfo{author}{Malnou, M.} \emph{et~al.}
\newblock \bibinfo{title}{Squeezed vacuum used to accelerate the search for a
  weak classical signal}.
\newblock \emph{\bibinfo{journal}{Phys. Rev. X}} \textbf{\bibinfo{volume}{9}},
  \bibinfo{pages}{021023} (\bibinfo{year}{2019}).

\bibitem{buschmann2020sims}
\bibinfo{author}{Buschmann, M.}, \bibinfo{author}{Foster, J.~W.} \&
  \bibinfo{author}{Safdi, B.~R.}
\newblock \bibinfo{title}{Early-universe simulations of the cosmological
  axion}.
\newblock \emph{\bibinfo{journal}{Phys. Rev. Lett.}}
  \textbf{\bibinfo{volume}{124}}, \bibinfo{pages}{161103}
  (\bibinfo{year}{2020}).

\bibitem{klaer2017mass26}
\bibinfo{author}{Klaer, V.~B.} \& \bibinfo{author}{Moore, G.~D.}
\newblock \bibinfo{title}{The dark-matter axion mass}.
\newblock \emph{\bibinfo{journal}{J. Cosmol. Astropart. Phys.}}
  \textbf{\bibinfo{volume}{2017}}, \bibinfo{pages}{049--049}
  (\bibinfo{year}{2017}).

\bibitem{slusher1985observation}
\bibinfo{author}{Slusher, R.~E.}, \bibinfo{author}{Hollberg, L.~W.},
  \bibinfo{author}{Yurke, B.}, \bibinfo{author}{Mertz, J.~C.} \&
  \bibinfo{author}{Valley, J.~F.}
\newblock \bibinfo{title}{Observation of squeezed states generated by four-wave
  mixing in an optical cavity}.
\newblock \emph{\bibinfo{journal}{Phys. Rev. Lett.}}
  \textbf{\bibinfo{volume}{55}}, \bibinfo{pages}{2409--2412}
  (\bibinfo{year}{1985}).

\bibitem{tse2019quantum}
\bibinfo{author}{Tse, M.} \emph{et~al.}
\newblock \bibinfo{title}{{Quantum-Enhanced Advanced LIGO Detectors in the Era
  of Gravitational-Wave Astronomy}}.
\newblock \emph{\bibinfo{journal}{Phys. Rev. Lett.}}
  \textbf{\bibinfo{volume}{123}}, \bibinfo{pages}{231107}
  (\bibinfo{year}{2019}).

\bibitem{brubaker2017first}
\bibinfo{author}{Brubaker, B.~M.} \emph{et~al.}
\newblock \bibinfo{title}{First results from a microwave cavity axion search at
  $24\text{ }\text{ }\ensuremath{\mu}\mathrm{eV}$}.
\newblock \emph{\bibinfo{journal}{Phys. Rev. Lett.}}
  \textbf{\bibinfo{volume}{118}}, \bibinfo{pages}{061302}
  (\bibinfo{year}{2017}).

\bibitem{peccei1977CP}
\bibinfo{author}{Peccei, R.~D.} \& \bibinfo{author}{Quinn, H.~R.}
\newblock \bibinfo{title}{{$CP$} conservation in the presence of
  pseudoparticles}.
\newblock \emph{\bibinfo{journal}{Phys. Rev. Lett.}}
  \textbf{\bibinfo{volume}{38}}, \bibinfo{pages}{1440} (\bibinfo{year}{1977}).

\bibitem{preskill1983cosmology}
\bibinfo{author}{Preskill, J.}, \bibinfo{author}{Wise, M.~B.} \&
  \bibinfo{author}{Wilczek, F.}
\newblock \bibinfo{title}{Cosmology of the invisible axion}.
\newblock \emph{\bibinfo{journal}{Phys. Lett. B}}
  \textbf{\bibinfo{volume}{120}}, \bibinfo{pages}{127} (\bibinfo{year}{1983}).

\bibitem{dine1983harmless}
\bibinfo{author}{Dine, M.} \& \bibinfo{author}{Fischler, W.}
\newblock \bibinfo{title}{The not-so-harmless axion}.
\newblock \emph{\bibinfo{journal}{Phys. Lett. B}}
  \textbf{\bibinfo{volume}{120}}, \bibinfo{pages}{137--141}
  (\bibinfo{year}{1983}).

\bibitem{Abbott1983bound}
\bibinfo{author}{Abbott, L.} \& \bibinfo{author}{Sikivie, P.}
\newblock \bibinfo{title}{{A Cosmological Bound on the Invisible Axion}}.
\newblock \emph{\bibinfo{journal}{Phys. Lett. B}}
  \textbf{\bibinfo{volume}{120}}, \bibinfo{pages}{133--136}
  (\bibinfo{year}{1983}).

\bibitem{ade2016planck}
\bibinfo{author}{Ade, P.~A.} \emph{et~al.}
\newblock \bibinfo{title}{Planck 2015 results-{XIII}. {C}osmological
  parameters}.
\newblock \emph{\bibinfo{journal}{A\&A}} \textbf{\bibinfo{volume}{594}},
  \bibinfo{pages}{A13} (\bibinfo{year}{2016}).

\bibitem{ouellet2019first}
\bibinfo{author}{Ouellet, J.~L.} \emph{et~al.}
\newblock \bibinfo{title}{First results from {ABRACADABRA}-10 cm: A search for
  sub-$\mu$ev axion dark matter}.
\newblock \emph{\bibinfo{journal}{Phys. Rev. Lett.}}
  \textbf{\bibinfo{volume}{122}}, \bibinfo{pages}{121802}
  (\bibinfo{year}{2019}).

\bibitem{majorovits2020madmax}
\bibinfo{author}{Majorovits, B.} \emph{et~al.}
\newblock \bibinfo{title}{Madmax: A new road to axion dark matter detection}.
\newblock In \emph{\bibinfo{booktitle}{J. Phys. Conf. Ser.}}, vol.
  \bibinfo{volume}{1342}, \bibinfo{pages}{012098} (\bibinfo{organization}{IOP
  Publishing}, \bibinfo{year}{2020}).

\bibitem{Geraci2014ARIADNE}
\bibinfo{author}{Arvanitaki, A.} \& \bibinfo{author}{Geraci, A.~A.}
\newblock \bibinfo{title}{Resonantly detecting axion-mediated forces with
  nuclear magnetic resonance}.
\newblock \emph{\bibinfo{journal}{Phys. Rev. Lett.}}
  \textbf{\bibinfo{volume}{113}}, \bibinfo{pages}{161801}
  (\bibinfo{year}{2014}).

\bibitem{Garcon2017CASPr}
\bibinfo{author}{Garcon, A.} \emph{et~al.}
\newblock \bibinfo{title}{The cosmic axion spin precession experiment
  ({CASPEr}): a dark-matter search with nuclear magnetic resonance}.
\newblock \emph{\bibinfo{journal}{Quantum Sci. Technol.}}
  \textbf{\bibinfo{volume}{3}}, \bibinfo{pages}{014008} (\bibinfo{year}{2017}).

\bibitem{malnou2018optimal}
\bibinfo{author}{Malnou, M.}, \bibinfo{author}{Palken, D.~A.},
  \bibinfo{author}{Vale, L.~R.}, \bibinfo{author}{Hilton, G.~C.} \&
  \bibinfo{author}{Lehnert, K.~W.}
\newblock \bibinfo{title}{Optimal operation of a {Josephson} parametric
  amplifier for vacuum squeezing}.
\newblock \emph{\bibinfo{journal}{Phys. Rev. Appl.}}
  \textbf{\bibinfo{volume}{9}}, \bibinfo{pages}{044023} (\bibinfo{year}{2018}).

\bibitem{Bertone2018WIMP}
\bibinfo{author}{Bertone, G.} \& \bibinfo{author}{Tait, T. M.~P.}
\newblock \bibinfo{title}{A new era in the search for dark matter}.
\newblock \emph{\bibinfo{journal}{Nature}} \textbf{\bibinfo{volume}{562}},
  \bibinfo{pages}{51--56} (\bibinfo{year}{2018}).

\bibitem{zhong2018results}
\bibinfo{author}{Zhong, L.} \emph{et~al.}
\newblock \bibinfo{title}{Results from phase 1 of the {HAYSTAC} microwave
  cavity axion experiment}.
\newblock \emph{\bibinfo{journal}{Phys. Rev. D}} \textbf{\bibinfo{volume}{97}},
  \bibinfo{pages}{092001} (\bibinfo{year}{2018}).

\bibitem{braine2020extended_ADMX}
\bibinfo{author}{Braine, T.} \emph{et~al.}
\newblock \bibinfo{title}{Extended search for the invisible axion with the
  axion dark matter experiment}.
\newblock \emph{\bibinfo{journal}{Phys. Rev. Lett.}}
  \textbf{\bibinfo{volume}{124}}, \bibinfo{pages}{101303}
  (\bibinfo{year}{2020}).

\bibitem{lee2020CAPP}
\bibinfo{author}{Lee, S.}, \bibinfo{author}{Ahn, S.}, \bibinfo{author}{Choi,
  J.}, \bibinfo{author}{Ko, B.~R.} \& \bibinfo{author}{Semertzidis, Y.~K.}
\newblock \bibinfo{title}{Axion dark matter search around $6.7\text{
  }\ensuremath{\mu}\mathrm{eV}$}.
\newblock \emph{\bibinfo{journal}{Phys. Rev. Lett.}}
  \textbf{\bibinfo{volume}{124}}, \bibinfo{pages}{101802}
  (\bibinfo{year}{2020}).

\bibitem{sikivie1983haloscope}
\bibinfo{author}{Sikivie, P.}
\newblock \bibinfo{title}{Experimental tests of the ``invisible" axion}.
\newblock \emph{\bibinfo{journal}{Phys. Rev. Lett.}}
  \textbf{\bibinfo{volume}{51}}, \bibinfo{pages}{1415--1417}
  (\bibinfo{year}{1983}).

\bibitem{Braunstein2005wigner}
\bibinfo{author}{Braunstein, S.~L.} \& \bibinfo{author}{van Loock, P.}
\newblock \bibinfo{title}{Quantum information with continuous variables}.
\newblock \emph{\bibinfo{journal}{Rev. Mod. Phys.}}
  \textbf{\bibinfo{volume}{77}}, \bibinfo{pages}{513--577}
  (\bibinfo{year}{2005}).

\bibitem{CavesThorne1982oscillator}
\bibinfo{author}{Caves, C.~M.}, \bibinfo{author}{Thorne, K.~S.},
  \bibinfo{author}{Drever, R. W.~P.}, \bibinfo{author}{Sandberg, V.~D.} \&
  \bibinfo{author}{Zimmermann, M.}
\newblock \bibinfo{title}{On the measurement of a weak classical force coupled
  to a quantum-mechanical oscillator. {I}. {I}ssues of principle}.
\newblock \emph{\bibinfo{journal}{Rev. Mod. Phys.}}
  \textbf{\bibinfo{volume}{52}}, \bibinfo{pages}{341--392}
  (\bibinfo{year}{1980}).

\bibitem{palken2020improved}
\bibinfo{author}{Palken, D.~A.} \emph{et~al.}
\newblock \bibinfo{title}{Improved analysis framework for axion dark matter
  searches}.
\newblock \emph{\bibinfo{journal}{Phys. Rev. D}}
  \textbf{\bibinfo{volume}{101}}, \bibinfo{pages}{123011}
  (\bibinfo{year}{2020}).

\bibitem{kim1979KSVZ}
\bibinfo{author}{Kim, J.~E.}
\newblock \bibinfo{title}{Weak-{Interaction} {Singlet} and {Strong} {$CP$}
  {Invariance}}.
\newblock \emph{\bibinfo{journal}{Phys. Rev. Lett.}}
  \textbf{\bibinfo{volume}{43}}, \bibinfo{pages}{103--107}
  (\bibinfo{year}{1979}).

\bibitem{shifman1980KSVZ2}
\bibinfo{author}{Shifman, M.~A.}, \bibinfo{author}{Vainshtein, A.~I.} \&
  \bibinfo{author}{Zakharov, V.~I.}
\newblock \bibinfo{title}{Can confinement ensure natural {$CP$} invariance of
  strong interactions?}
\newblock \emph{\bibinfo{journal}{Nucl. Phys. B}}
  \textbf{\bibinfo{volume}{166}}, \bibinfo{pages}{493--506}
  (\bibinfo{year}{1980}).

\bibitem{yamamoto2008flux}
\bibinfo{author}{Yamamoto, T.} \emph{et~al.}
\newblock \bibinfo{title}{Flux-driven {J}osephson parametric amplifier}.
\newblock \emph{\bibinfo{journal}{Appl. Phys. Lett.}}
  \textbf{\bibinfo{volume}{93}}, \bibinfo{pages}{042510}
  (\bibinfo{year}{2008}).

\bibitem{caves1982quantum}
\bibinfo{author}{Caves, C.~M.}
\newblock \bibinfo{title}{Quantum limits on noise in linear amplifiers}.
\newblock \emph{\bibinfo{journal}{Phys. Rev. D}} \textbf{\bibinfo{volume}{26}},
  \bibinfo{pages}{1817--1839} (\bibinfo{year}{1982}).

\bibitem{Rapidis2019cavity}
\bibinfo{author}{Rapidis, N.~M.}, \bibinfo{author}{Lewis, S.~M.} \&
  \bibinfo{author}{van Bibber, K.}
\newblock \bibinfo{title}{Characterization of the {HAYSTAC} axion dark matter
  search cavity using microwave measurement and simulation techniques}.
\newblock \emph{\bibinfo{journal}{Rev. Sci. Instrum.}}
  \textbf{\bibinfo{volume}{90}} (\bibinfo{year}{2019}).

\bibitem{brubaker2017analysis}
\bibinfo{author}{Brubaker, B.~M.}, \bibinfo{author}{Zhong, L.},
  \bibinfo{author}{Lamoreaux, S.~K.}, \bibinfo{author}{Lehnert, K.~W.} \&
  \bibinfo{author}{van Bibber, K.~A.}
\newblock \bibinfo{title}{{HAYSTAC} axion search analysis procedure}.
\newblock \emph{\bibinfo{journal}{Phys. Rev. D}} \textbf{\bibinfo{volume}{96}},
  \bibinfo{pages}{123008} (\bibinfo{year}{2017}).

\bibitem{Tanabashi2018PDB}
\bibinfo{author}{Tanabashi, M.} \emph{et~al.}
\newblock \bibinfo{title}{Review of particle physics}.
\newblock \emph{\bibinfo{journal}{Phys. Rev. D}} \textbf{\bibinfo{volume}{98}},
  \bibinfo{pages}{030001} (\bibinfo{year}{2018}).

\bibitem{cheng1995axion}
\bibinfo{author}{Cheng, S.~L.}, \bibinfo{author}{Geng, C.~Q.} \&
  \bibinfo{author}{Ni, W.-T.}
\newblock \bibinfo{title}{Axion-photon couplings in invisible axion models}.
\newblock \emph{\bibinfo{journal}{Phys. Rev. D}} \textbf{\bibinfo{volume}{52}},
  \bibinfo{pages}{3132--3135} (\bibinfo{year}{1995}).

\bibitem{dine1981DFSZ}
\bibinfo{author}{Dine, M.}, \bibinfo{author}{Fischler, W.} \&
  \bibinfo{author}{Srednicki, M.}
\newblock \bibinfo{title}{A simple solution to the strong {$CP$} problem with a
  harmless axion}.
\newblock \emph{\bibinfo{journal}{Phys. Lett. B}}
  \textbf{\bibinfo{volume}{104}}, \bibinfo{pages}{199--202}
  (\bibinfo{year}{1981}).

\bibitem{zhit1980DFSZ2}
\bibinfo{author}{Zhitnitsky, A.~R.}
\newblock \bibinfo{title}{{On Possible Suppression of the Axion Hadron
  Interactions. (In Russian)}}.
\newblock \emph{\bibinfo{journal}{Sov. J. Nucl. Phys.}}
  \textbf{\bibinfo{volume}{31}}, \bibinfo{pages}{260} (\bibinfo{year}{1980}).

\bibitem{burkhart2020error}
\bibinfo{author}{Burkhart, L.~D.} \emph{et~al.}
\newblock \bibinfo{title}{Error-detected state transfer and entanglement in a
  superconducting quantum network.} \bibinfo{note}{{a}rXiv preprint
  arxiv:2004.06168 (2020)}.

\bibitem{Kenany2017design}
\bibinfo{author}{Al~Kenany, S.} \emph{et~al.}
\newblock \bibinfo{title}{Design and operational experience of a microwave
  cavity axion detector for the 20 -- 100$\,\mu$ev range}.
\newblock \emph{\bibinfo{journal}{Nucl. Instrum. Methods Phys. Res.}}
  \textbf{\bibinfo{volume}{854}}, \bibinfo{pages}{11--24}
  (\bibinfo{year}{2017}).

\end{thebibliography}
%\end{multicols}{1}
\end{document}